\documentclass[]{aa}

\usepackage{graphicx}
\usepackage{txfonts}
\usepackage{natbib}
\usepackage{morefloats}
\bibpunct{(}{)}{;}{a}{}{,}

\begin{document} 

\title{Coupling hydrodynamics and radiation calculations for star-jet interactions in AGN}

\author{ 
V. M. de~la~Cita\inst{1}
\and V. Bosch-Ramon\inst{1}
\and X. Paredes-Fortuny\inst{1}
\and D. Khangulyan\inst{2}
\and M. Perucho\inst{3,4}
}

\institute{Departament d'Astronomia i Meteorologia, Institut de Ci\`ences del Cosmos (ICCUB),
Universitat de Barcelona (IEEC-UB), Mart\'{\i} i Franqu\`es 1,
E-08028 Barcelona, Spain
\and Department of Physics, Rikkyo University 3-34-1, Nishi-Ikebukuro, Toshima-ku, Tokyo 171-8501, Japan
\and Departament d'Astronomia i Astrof\'{\i}sica, Universitat de Val\`encia, 
Av.\ Vicent Andr\'es Estell\'es s/n, 46100 Burjassot (Val\`encia), Spain
\and Observatori Astron\`omic, Universitat de Val\`encia, C/ Catedr\`atic Jos\'e Beltran, 2, 46980, Paterna (Val\`encia), Spain 
}
\offprints{V. M. de~la~Cita, \email{vmdelacita@am.ub.es}}
   \date{Received 29 July 2015 ; accepted 15 April 2016}
 
  \abstract
{Stars and their winds can contribute to the non-thermal emission in extragalactic jets. Because of the complexity of jet-star interactions, the properties of the resulting emission are closely linked to those of the emitting flows.}
{We simulate the interaction between a stellar wind and a relativistic extragalactic jet and use the hydrodynamic results to compute the non-thermal emission under different conditions.}  
{We performed relativistic axisymmetric hydrodynamical simulations of a relativistic jet interacting with a supersonic, non-relativistic stellar wind. We computed the corresponding streamlines out of the simulation results and calculated the injection, evolution, and emission of non-thermal particles accelerated in the jet shock, focusing on electrons or $e^\pm$-pairs. Several cases were explored, considering different jet-star interaction locations, magnetic fields, and observer lines of sight. The jet luminosity and star properties were fixed, but the results are easily scalable when these parameters are changed.}    
{Individual jet-star interactions produce synchrotron and inverse Compton emission that peaks from X-rays to MeV energies (depending on the magnetic field), and at $\sim 100-1000$~GeV (depending on the stellar type), respectively. The radiation spectrum is hard in the scenarios explored here as a result of non-radiative cooling dominance, as low-energy electrons are efficiently advected even under relatively high magnetic fields. Interactions of jets with cold stars lead to an even harder inverse Compton spectrum because of the Klein-Nishina effect in the cross section. Doppler boosting has a strong effect on the observer luminosity.}
{The emission levels for individual interactions found here are in the line of previous, more approximate, estimates, strengthening the hypothesis that collective jet-star interactions could significantly contribute at high energies under efficient particle acceleration.} 


   \maketitle

\section{Introduction}

The winds and atmospheres of stars have been proposed to play an important role in the propagation, matter content, stability, and potential disruption of the jets of active galactic nuclei (AGN) \citep[e.g.][]{Kom94,blk96,HubBla06,bpb12,pml14}. In addition, the interaction of jets with stellar atmospheres or winds have also been suggested to lead to non-thermal emission that may be detectable from Earth, in both blazar and non-blazar AGN, and in the form of both transient and persistent radiation \citep[e.g.][]{BedPro97,bab10,bab12,bba12,kbb13,bpb12,abr13,Bos15,BedBan15}. It is noteworthy that there might already be direct observational evidence of such an interaction \citep[e.g.][]{hwk03,mko14}, and that high-energy phenomena observed in some AGN might be interpreted in the context of jet-star interactions \citep[e.g.][]{bab10,bab12,bba12,kbb13}. The actual extent of the dynamical and radiative impact of these interactions is still unknown, however.

In the jet-star interaction scenario, stars are expected to cross AGN jets at different distances from the jet base, producing shocks that can transfer kinetic energy to non-thermal particles. In the case of highly magnetized jets, these would present different mechanisms to transfer jet energy, now in the magnetic field instead of being in kinetic form, to non-thermal particles \citep[see, e.g.][and references therein]{Bos12}. Electric fields produced by jet-stretched magnetic field lines around an obstacle would also accelerate particles \citep[e.g.][]{jrt96}.

In the  inner-most jet regions, stars are expected to move fast and the jet is narrower, which means that only a few stars can interact with the jet at a time. Here  the emission might be released during relatively short events that are triggered by high-inertia targets, such as the external weakly bound layers of evolved giants \citep[e.g.][]{bab10}, or stars with very high mass-loss \citep[e.g.][]{abr13}. Persistent emission might also take place far from the jet base, as the jet propagates through the inner-most kpc regions of the
galaxy, and even farther out \citep[][]{abr13,BedBan15}. In all these situations, several
ingredients are required to accurately estimate the interaction duration, rate, effect on the jet properties, and related radiation: a proper characterization of the stellar populations and their spatial distribution, both galaxy-type dependent, and a detailed description of the physics of the jet-star interaction and the associated non-thermal processes. An approximate study combining both hydrodynamics and radiation estimates was carried out, the
result of which was that the emission from individual jet-star interactions may be significantly higher than previously thought \citep{Bos15}.

A proper understanding of the AGN jet radiation and its underlying physics needs an accurate characterization of the emission produced in stars interacting with AGN jets. 
To proceed in this direction, following previous works, we here combine relativistic
hydrodynamical (RHD) simulations and radiation calculations, assuming no dynamical feedback from non-thermal processes (a rather good assumption in the cases studied here), to characterize the emission produced in the shocked jet region that forms when the jet flow is stopped by a stellar
wind.
First we simulate the interaction of a stellar wind and a relativistic jet until steady-state is reached. Then, the obtained hydrodynamical information is used to characterize the injection of non-thermal particles, their
propagation, and emission. As matter and radiation densities are generally low unless interactions occur close to the jet base \citep[see][]{bab12,kbb13}, hadronic processes are not expected to be relevant. Thus, the emitting particles are assumed to be leptons here, which might be electrons for a proton-dominated jet, or electrons and positrons ($e^\pm$) for a $e^\pm$-dominated jet. We consider a situation in which a star of relatively high mass-loss and luminosity, such as a red giant or a moderately early star, interacts with a jet of intermediate power. The results can be scaled, which allows deriving broader conclusions. Throughout this paper, primed quantities are in the fluid reference frame (FF).

\section{Hydrodynamics}\label{sect:hydro}

\subsection{Simulated cases and streamline preparation}

We performed axisymmetric RHD simulations of the interaction between a relativistic jet and a stellar wind within the jet. The simulations were conducted using a finite-difference code based on a high-resolution shock-capturing scheme that solves the equations of RHD in two dimensions (2D) in a conservation form \citep{mmf97}. The code is parallelized using open message passing (OpenMP; \citealt{pmh05}). The simulations were run in a workstation with two Intel(R) Xeon(R) CPU E5-2643 processors (3.30 GHz, $4\times2$ cores, with two threads for each core) and four modules of 4096 MB of memory (DDR3 at 1600 MHz). The obtention of streamlines from the simulation results allowed characterizing the regions of interest as many 1D hydrodynamical structures, which is suitable for radiation calculations.

\subsubsection{Jet-star interaction}\label{sec:simu_jet-star}

We assumed a collisionless adiabatic and relativistic ideal gas with a dynamically negligible magnetic field. For simplicity, the gas has one particle species with a constant relativistic adiabatic index $\hat\gamma$ of 4/3 for both the jet and the stellar wind material. It is not required for our purposes at this point to determine whether the jet is made of protons (nuclei) plus electrons, or $e^\pm$-pairs.
The physical size of the domain is $r \in \lbrack 0, l_r \rbrack$ with $l_r = 2\times10^{15}$~cm, and $z \in \lbrack 0, l_z \rbrack$ with $l_z = 1.5\times10^{15}$~cm. The total number of cells is 400 and 300 in the radial and axial directions, respectively. This resolution was chosen to have enough numerical dissipation to avoid a growth of instabilities that is fast enough to prevent the formation of a quasi-steady state (see Sect.~\ref{results}). The star was located at $(r_0, z_0) = (0, 0.3\times10^{15})$~cm, and its spherical wind was injected through a region with radius $r_{\rm in} = 7\times10^{13}$~cm (14 cells), small enough not to be affected by the shock terminating this wind. The jet was injected at the bottom boundary of the grid. The jet streamlines were approximated as parallel instead of radially extending from the jet origin because the scales of the simulation were much smaller than the height of the jet at which the interaction takes place. The upper and right boundaries of the grid were set to outflow, while the left boundary was set to reflection. All these parameters are summarized in Table~\ref{table_sim}. 

\begin{table}
\caption{Simulation parameters.}
\label{table_sim}
\centering
\begin{tabular}{c c}
\hline\hline
Parameter & Jet-stellar wind simulation \\
\hline
$\hat\gamma$     & $4/3$ \\
$l_r$        & $2\times10^{15}~\rm{cm}$ \\
$l_z$        & $1.5\times10^{15}~\rm{cm}$ \\
$n_r$        & $400$ \\       
$n_z$        & $300$ \\       
\hline
\end{tabular}
\tablefoot{Adiabatic index $\hat\gamma$, 
physical $r$-grid size $l_r$, physical $z$-grid size $l_z$, 
number of cells in the $r$-axis $n_r$, and number of cells in the $z$-axis $n_z$.}
\end{table}

The physical parameters of the jet, which is in an inflow condition at the bottom of the computational grid, are the total jet power within the grid, $L_{\rm 0}\approx 4\times 10^{37}~{\rm erg~s^{-1}}$ ($\sim 10^{44}$~erg~s$^{-1}$ for a 1~pc jet radius), the Lorentz factor $\Gamma_{\rm 0} = 1/\sqrt{1-(v_{\rm 0}/c)^2}=10$, with $v_{\rm 0}=v_z=0.995~c$ and $v_r=0$, and the specific internal energy $\epsilon_{\rm 0} = 9\times10^{18}~{\rm erg~g^{-1}}$. The jet power is computed as\begin{equation}
L_0 = \pi\,l_r^2\,\Gamma_{\rm 0}\rho_0(h_0\Gamma_{\rm 0} -1)c^2v_0\,,
\end{equation}
where $\rho_{\rm 0} = 1.24\times 10^{-27}~{\rm g~cm^{-3}}$ is the jet density, $p_{\rm 0} = \left(\hat\gamma-1\right)\rho_{\rm 0}\epsilon_{\rm 0}$ its pressure, and $h_{\rm 0}=1+\frac{\epsilon_{\rm 0}}{c^2}+\frac{p_{\rm 0}}{\rho_{\rm 0} {c^2}}$ its specific enthalpy.

The stellar wind is a spherical inflow condition imposed at $7\times10^{13}$~cm from the star centre. The wind physical parameters at injection are the mass-loss rate $\dot{M}=10^{-9}~{\rm M_\odot~yr^{-1}}$, the radial velocity $v_{\rm sw}= 2\times 10^8$~cm~s$^{-1}$, and the specific internal energy $\epsilon_{\rm sw} = 9\times10^{13}~{\rm erg~g^{-1}}$. The derived stellar wind density at injection is $\rho_{\rm sw}\approx 5.2\times10^{-21}~{\rm g~cm^{-3}}$. The stellar wind is taken to be homogeneous, meaning that it is not clumpy, and since it is supersonic, its density profile from the injection radius up to its termination is $\propto 1/R^2$, with $R$ being the distance to the star centre. The wind properties correspond to those of a high-mass star with a modest mass-loss rate. The thrust of this wind, $\dot{M}v_{\rm sw}\approx 1.3\times 10^{25}$~g~cm~s$^{-2}$, would correspond to that of a red giant, although if a red giant wind had been simulated, the velocity would have been an order of magnitude lower (with $\dot{M}$ scaling accordingly), making the simulation much longer. However, since the relation between the jet momentum flux and the wind thrust determines to first order the shape of the interaction region, a lighter wind of equal thrust can be used to reduce the computational costs.

The star was assumed to be at rest. This is a reasonable assumption as long as the stellar velocity is much lower than $v_{\rm sw}$. Otherwise, the jet-wind interaction geometry will strongly depart from axisymmetry, making the results obtained here less realistic. The Keplerian velocity for a $10^8$~M$_\odot$ central black hole at a distance of 10~pc is $v_{\rm K}\approx 2\times 10^7$~cm~s$^{-1}$, a 10\% of the adopted $v_{\rm sw}$-value. We considered this $v_{\rm K}$-value low enough at this stage, but a caveat must be made: for a more realistic red giant wind, $v_{\rm K}$ would become of the order of $v_{\rm sw}$, and the star motion would then have to be taken into account. This requires 3D simulations,
however, which are much more computationally expensive than the axisymmetric simulations carried out here, which are meant as a first step in coupling radiation and hydrodynamics studying jet-star interactions (we note, nevertheless, that work is being carried out in this direction). Finally we note that in general the jet-crossing time will be much longer than the simulated times (given in Sect.~\ref{results}).

The grid was initially filled with the jet properties except within the stellar wind
injection region, which was filled with the stellar wind properties. The jet and stellar wind physical parameters are
summarized in Table~\ref{table_wind}.

\begin{table}
\caption{Jet and stellar wind physical parameters.}
\label{table_wind}
\centering
\begin{tabular}{c c c}
\hline\hline
Parameter    & Jet                                  & Stellar wind                         \\
\hline                                                                                    
$\rho$       & $1.24\times10^{-27}~{\rm g~cm^{-3}}$  & $5.12\times10^{-21}~{\rm g~cm^{-3}}$  \\
$\epsilon$   & $9\times10^{18}~{\rm erg~g^{-1}}$   & $9\times10^{13}~{\rm erg~g^{-1}}$   \\
$v$          & $0.995~{c}$                         & $2000~{\rm km~s^{-1}}$               \\
\hline
\end{tabular}
\tablefoot{
Density $\rho$, specific internal energy $\epsilon$, and velocity $v$
at $z=1\times10^{13}$~cm for the jet (boundary condition), 
at a distance $r_{\rm in} \le 7\times10^{13}$~cm with respect to the star centre at $(r_0,z_0) = (0, 0.3\times10^{15})$~cm (boundary condition).
}
\end{table}

The jet momentum flux (thermal plus kinetic pressure) at the bottom of the computational grid is
\begin{equation}
F_{\rm 0}=\rho_{\rm 0} \Gamma_{\rm 0}^2 v_{\rm 0}^2 h_{\rm 0}+p_{\rm 0}\,{\rm ,}
\label{eta1}
\end{equation}
and the stellar wind momentum flux at a distance $R$ is\begin{equation}
F_{\rm sw}=\frac{\dot{M}v_{\rm sw}}{S_{\rm sw}}+p_{\rm sw}\,{\rm ,}
\label{eta2}
\end{equation}
where $S_{\rm sw}=4\pi R^2$, and $p_{\rm sw} = \left(\hat\gamma-1\right)\rho_{\rm sw}\epsilon_{\rm sw}$ is the pressure of the stellar wind.
Equations (\ref{eta1}) and (\ref{eta2}) allow us to set the point where the jet and the stellar wind momentum fluxes are equal, at $R_{\rm CD}=0.1\times 10^{15}$~cm from the star centre, locating the contact discontinuity (CD) on the simulation axis. This corresponds to a position
$z_{\rm CD}=z_0-R_{\rm CD} = 0.2\times10^{15}~{\rm cm}$ with respect to the bottom of the grid ($z=0$).

\subsubsection{Streamline calculations}\label{strcal}

\begin{figure}
\centering
\resizebox{\hsize}{!}{\includegraphics{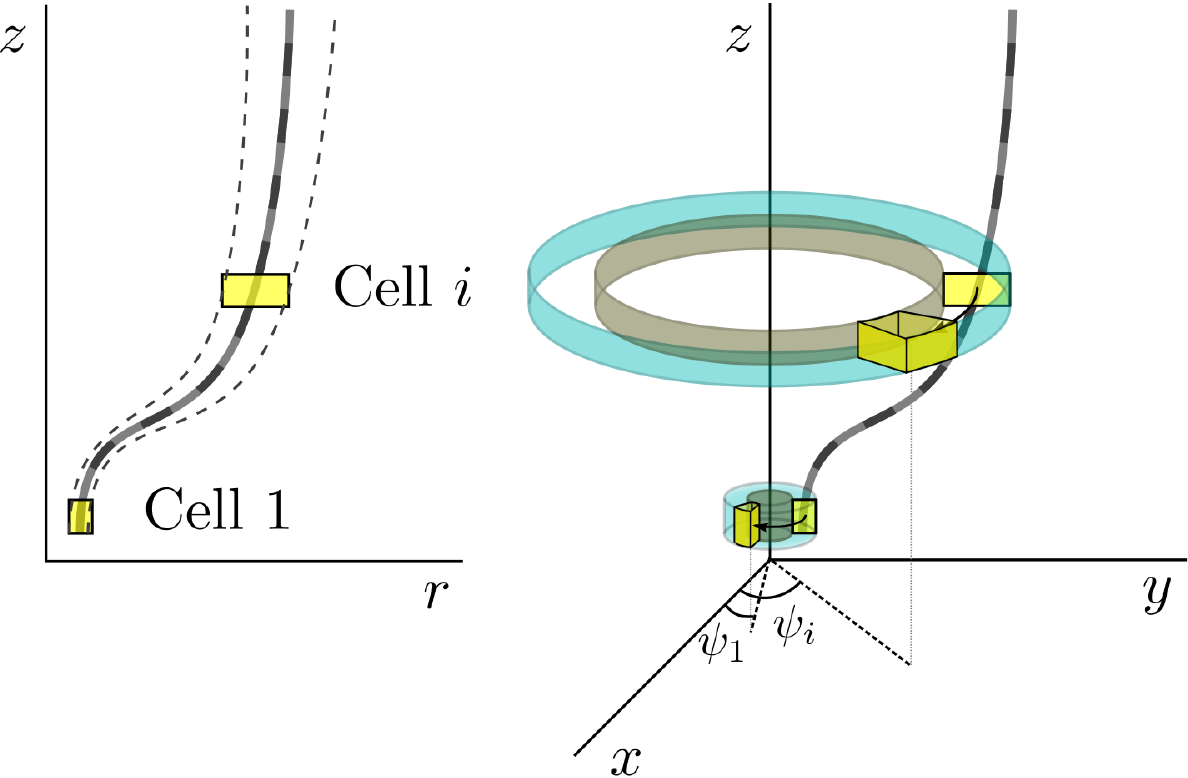}}
\caption{
Two cells in the axisymmetric representation (left) and in the real 3D space (right). In the latter, the annular structure that each cell represents is shown together with a picture of the assignment of the azimuthal angle $\psi_i$ .
}
\label{fig:scramble}
\end{figure}

Streamlines of the jet flow were characterized in the numerical solution as described in Appendix~\ref{strlin}. An important parameter determining the streamline magnetic field is $\chi_B$, the initial ratio of the Poynting-to-matter energy flux. The streamlines were divided into segments or cells. Each cell represents an annular element from the point of view of their volume and the determination of the non-thermal energy budget because the simulation is axisymmetric. On the other hand,  the cells represent point-like non-thermal
emitters in the radiation calculations below. To provide the 3D information of the structure of the whole non-thermal emitter, an azimuthal angle $\psi$ was assigned to each cell, as shown in Fig.\ref{fig:scramble}, and the value of this angle was varied from 0 to $2\pi$ to cover the whole emitting volume. Figure~\ref{fig:scramble} provides a sketch of the cell geometry in both contexts, the simulation results, and the radiation calculations.

\subsection{Results}
\label{results}

\subsubsection{Steady-state}

\begin{figure}
\centering
\resizebox{\hsize}{!}{\includegraphics{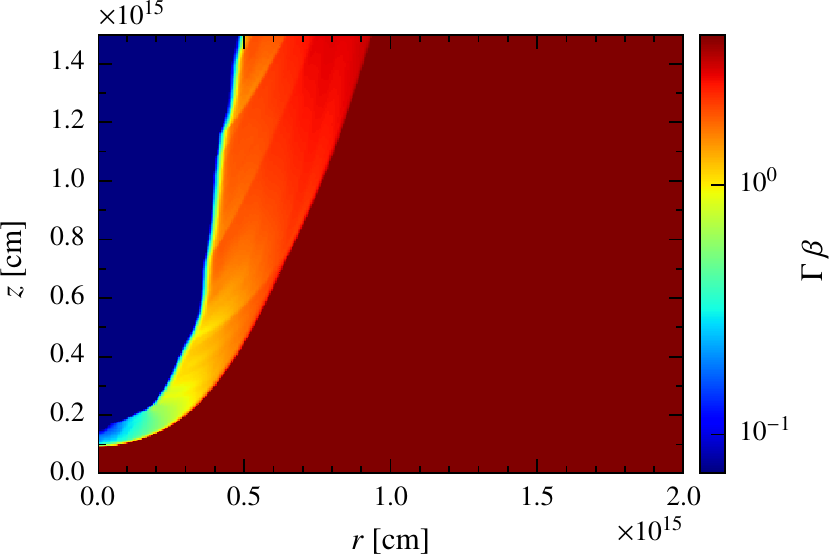}}
\caption{Distribution by colour of the module of the spatial component 
of the four-velocity at time $t = 3.3\times10^{7}$~s for the star-jet interaction in its steady configuration.}
\label{star_gamma}
\end{figure}

\begin{figure}
\centering
\resizebox{\hsize}{!}{\includegraphics{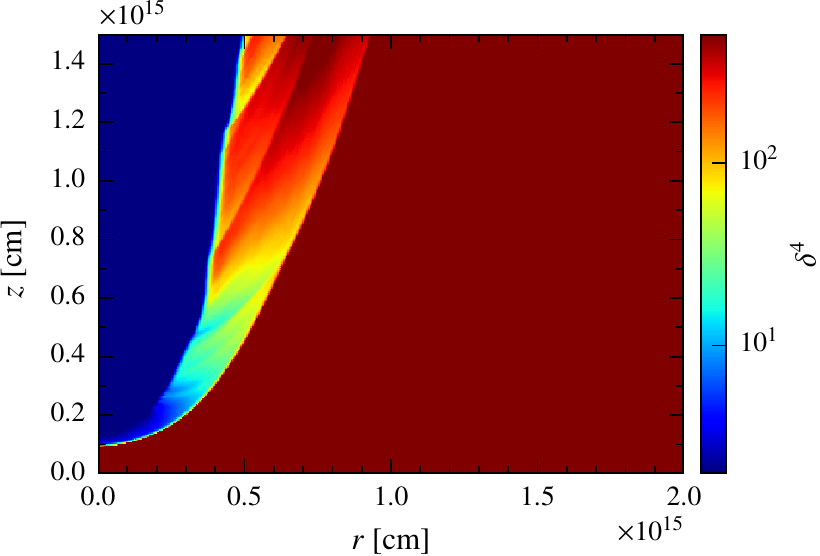}}
\caption{Distribution by colour of the Doppler-boosting enhancement of the emission, as seen from the top, at time $t = 3.3\times10^{7}$~s for the star-jet interaction in its steady configuration.}
\label{star_doppler}
\end{figure}

Figure~\ref{star_density_lines} shows the density map and the trajectories of the computed streamlines for the jet-stellar wind simulation when (quasi-) steady-state is reached at $t = 3.3\times10^{7}$~s. The dynamical timescale of the problem is determined by the stellar wind, which is $\sim l_z/v_{\rm sw}=7.5\times 10^6$~s, as it carries more mass and $v_{\rm sw}\ll v_0$. Thus, the $t$-value reached seems appropriate. The two shocks formed are bow-shaped towards the star, which has a much lower thrust than the jet within the grid. The CD in the simulation is at $z_{\rm CD}^{\rm num}\sim 0.2\times10^{15}$~cm, as obtained analytically in Sect.~\ref{sec:simu_jet-star}. Figure~\ref{star_gamma} illustrates the re-acceleration of the shocked jet material as it is advected upwards. This feature has a strong effect on the radiation because Doppler-boosting effects cannot be neglected. This is also shown in Fig.~\ref{star_doppler}, where the Doppler-boosting enhancement of the emission in the jet direction is presented. 

\subsubsection{Instability growth}

The simulation reaches a quasi-stationary numerical solution, with the shocked flow structure in a metastable state. There are recurrent perturbations coming from the numerical, spatial, and temporal discretization that grow as a result of the developing instabilities. Although the perturbations are of numerical origin, they can be considered to mimic the irregularities expected in real flows because they are hardly completely smooth or laminar. The double-shock structure presents variations in time caused by irregularities originated in the CD that grow as they are advected with the flow. The growth of these irregularities is mostly linked to the Kelvin-Helmholtz instability (KHI) because the velocity of the shocked jet flow along the CD grows very quickly from the simulation axis, which leads to a strong velocity difference with respect to the shocked wind; if the velocity of the CD perpendicular to itself were not zero, the Rayleigh-Taylor instability (RTI) would develop as well \citep{bookChandrasekhar}. In our case, the average velocity of the CD was zero (although there are small fluctuations in its position). The KHI is thus the dominant source of perturbation growth in the CD close and far from the axis.

The development of instabilities has imposed a limitation on the simulation parameters.  At the onset of the simulation, instabilities develop in the interacting flows, with the related growing irregularities being advected out of the grid and leaving the quasi-steady interaction structure described above. For lighter jets, heavier winds, or a higher resolution, the development and growth of the perturbations are enhanced and produce smaller and denser fragments of stellar wind able to penetrate deeper into the jet flow and generate shocks in the entire computational domain.  This is caused by instability growth, which causes small wind structures to quickly develop and propagate within the grid, triggering shocks in the jet flow. This can lead to a grid that is partially occupied by hot material that reaches the grid boundaries with subsonic velocities. This renders the simulation results unrealistic because waves develop in the grid boundaries and bounce back, which affects the flow dynamics inside the grid. In addition, since the flow is subsonic at the grid boundaries, the subsonic flow is evacuated too slowly and accumulates and can potentially end up filling the whole grid. This could be avoided by a larger grid, although if a higher resolution were used, the disruptive effect of instabilities \citep[see e.g.][]{phm04} would be enhanced, pushing the grid size requirements even further. 

Strong sensitivity to resolution and flow density contrast was already faced in previous similar axisymmetric simulations. The fast growth of perturbations close to the axis was seen for instance in \cite{pbp15} in the context of a relativistic pulsar wind interacting with a non-relativistic stellar wind. It was noted that this effect might have a partially numerical origin in the coordinate singularity plus a reflective boundary at $r=0$. However, a similarly fast perturbation growth also appeared in relativistic 2D simulations for that scenario \citep{bbk12,lfd13} in planar geometry, which shows that the perturbations were not only due to an artefact of the conditions at $r=0$. For the jet-stellar wind scenario, we ran low-resolution 3D simulations (not presented here) focusing on the region close to the star. We found that the shocked structure was also prone to develop instabilities, although the small grid size prevented a deeper analysis. 

\subsubsection{Perturbed state}

Judging from the discussion above,  this growth seems to be a physical effect to a large extent, although the fast growth of
the instabilities may be partially linked to the axisymmetry
of the simulations. It is thus worth considering some instance of the interaction region when it is affected by a strong perturbation before reaching steady-state. Such an instance is shown in Fig.~\ref{star_pert_density_lines}, which presents the density map and trajectories of the computed streamlines at $t = 3.7\times10^{6}$~s. As seen in the figure, the perturbations deeply penetrate the jet, increasing the size of the interaction region and strongly modifying the streamline trajectories. This is an example of how important physical instability growth might be for the effective size of the stellar target the jet meets. 

In the corner of the map in Fig.~\ref{star_pert_density_lines} where the top and right boundaries join, a reflection shock is visible in a small region. This reflection shock is an example of the presence of subsonic flow at the boundaries. However, this shock only affects a minor region of the simulation, even accounting for the cylindrical symmetry, so we have allowed for the presence of this small artefact. At our level of resolution, our simulation did reach a physically realistic steady-state (with the mentioned small fluctuations in the CD). Although the hydrodynamical solution shown in Fig.~\ref{star_pert_density_lines} does not correspond to steady-state because radiation is computed from the shocked jet material with a dynamical timescale $\sim l_z/c\ll l_z/v_{\rm sw}$, we can assume that the flow is in a pseudo-steady state for emission computation purposes. 

When characterizing the non-thermal emitters through streamlines, we discarded the lines that presented mixing between jet and stellar wind fluids that exceeded 50\% in at least one cell of the streamline for the jet-star interactions in steady-state. The surface of any discarded streamline was added to the line
immediate adjacent along the radial coordinate. This compensates for the lack of lines close to the jet axis because the hydrodynamical conditions are relatively similar there. On the other hand, the streamlines of the jet-star interaction in the perturbed state were computed regardless of the level of mixing because many lines were numerically affected. For this case, we just removed the segments of the streamlines that were affected by a mixing level $>50$\%.

\begin{figure}
\centering
\resizebox{\hsize}{!}{\includegraphics{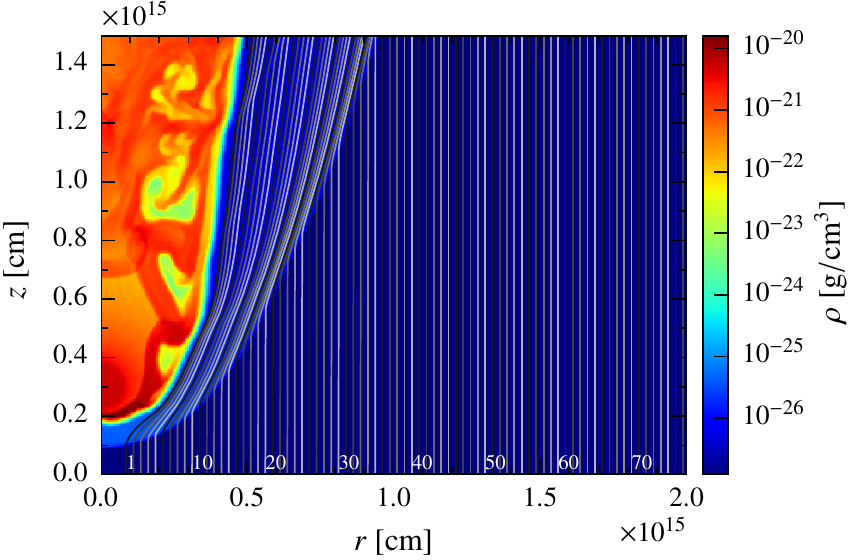}}
\caption{Density distribution by colour
at time $t = 3.3\times10^{7}$~s for the star-jet interaction in its steady configuration. 
The star is located at $(r_0, z_0) = (0, 3\times10^{14})$~cm and the jet is injected at $z=1\times10^{13}$~cm. The grey lines show the computed streamline trajectories; the numbers and the grey scale are added just for visualization.
}
\label{star_density_lines}
\end{figure}

\begin{figure}
\centering
\includegraphics{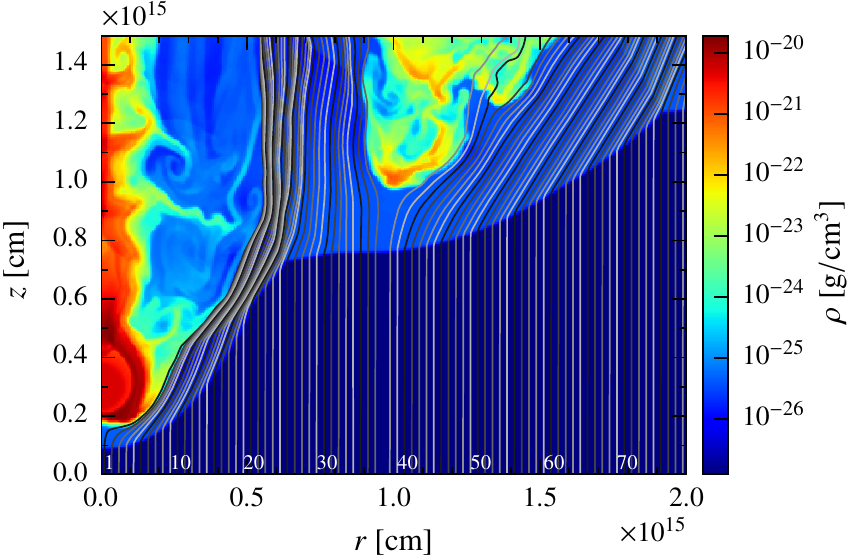}
\caption{Density distribution by colour at time $t = 3.7\times10^{6}$~s for the star-jet interaction in a perturbed state. The remaining plot properties are the same as those of Fig.~\ref{star_density_lines}.}
\label{star_pert_density_lines}
\end{figure}

\begin{figure}
\centering
\resizebox{\hsize}{!}{\includegraphics{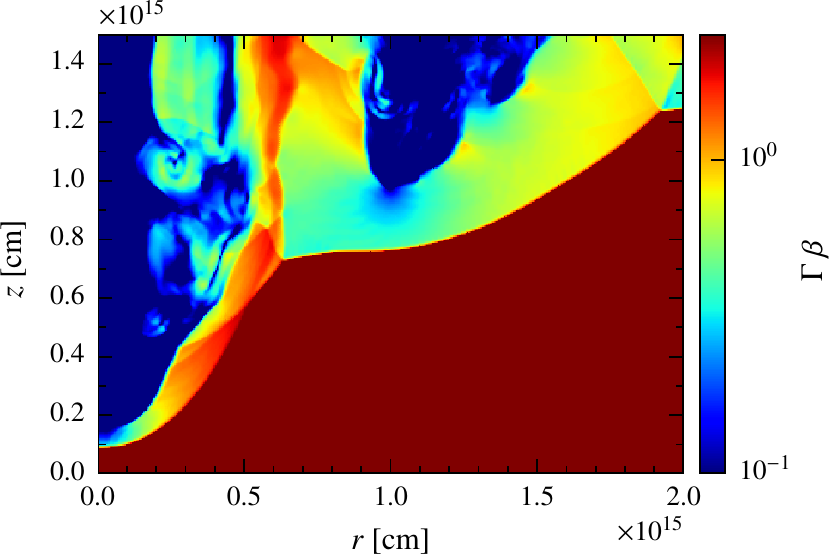}}
\caption{Distribution by colour of the module of the spatial component 
of the four-velocity at time $t = 3.7\times10^{6}$~s for the star-jet interaction in a perturbed state.}
\label{brac_gamma}
\end{figure}

\begin{figure}
\centering
\resizebox{\hsize}{!}{\includegraphics{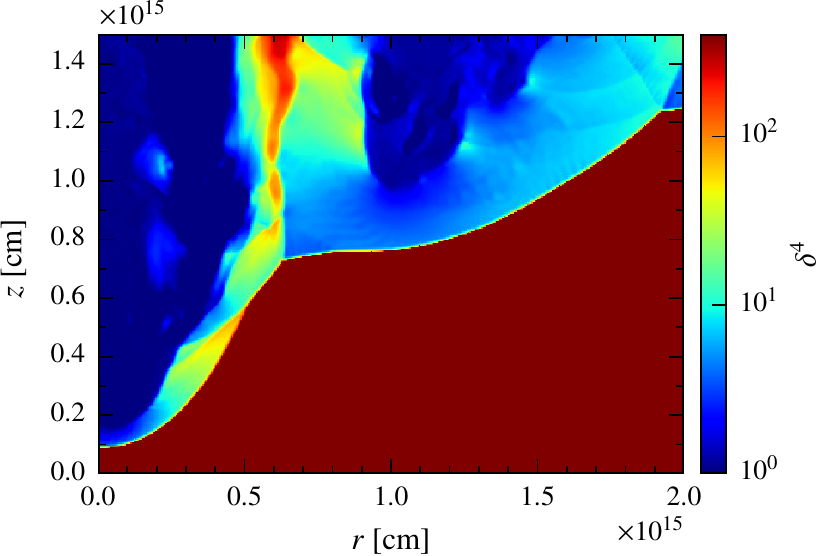}}
\caption{Distribution by colour of the Doppler boosting enhancement of the emission, as seen from the top, at time $t = 3.7\times10^{6}$~s, for the star-jet interaction in a perturbed state.}
\label{brac_doppler}
\end{figure}

\section{Radiation}

\subsection{Non-thermal emitter} \label{sec:radiative}

After obtaining the streamlines from the RHD simulations, we used the hydrodynamic information to compute the injection, cooling, and radiation of the non-thermal population. The computational domain was divided into cells as described in Sect.~\ref{strcal}. Each cell was considered homogeneous and characterized by its position and velocity vector information, pressure ($P$), density ($\rho$), section ($S$), magnetic field ($B$), and the flow velocity divergence ($\nabla(\Gamma\overrightarrow{v})$) to compute adiabatic losses. A parameter accounting for wind mixing, going from 0 ($100\%$ jet material) to 1 ($100\%$ stellar wind), was also taken into account through the use of a tracer variable that
is included in the RHD code.

 As described in Appendix~\ref{injpart}, the code computes for each streamline  (i) which cells have non-thermal particle injection
as well as the luminosity injected into these non-thermal particles, and (ii) the steady particle energy distribution in each cell, $N(E)$. The non-thermal particle injection was assumed to occur when the internal energy density and entropy grows, which is
the case for shocks in our hydrodynamical approach (in the case of resistive magnetohydrodynamical simulations, magnetic reconnection would also lead to an increase of the internal energy and entropy). Another important parameter is the fraction of energy in non-thermal particles, $\chi_\text{NT}$, which we fixed here to a modest 0.1, which was taken as a reference value. The radiation levels need to be scaled proportionally when higher or lower acceleration efficiencies are adopted. Note, however,  that high values of $\chi_{\rm NT}\lesssim 1$ would not be consistent with our hydrodynamical calculations if radiation cooling dominated (although in most of the instances computed in this work this is not the case).

After obtaining the particle energy distribution in each cell, as explained in Sect.~\ref{emission}, the code derives the synchrotron and inverse Compton (IC) radiation as seen from the observer. When the radiation coming from each line is known, the total spectral energy distribution (SED) can be obtained and emission maps derived. The code includes the possibility of accounting for time delays in the emission from non-stationary flows, but here this is not required as the emitting flow can be considered steady.

The hydrodynamic simulations are axisymmetric and 2D, but the actual emitter is a 3D structure. To properly compute Doppler boosting or obtain radiation maps for non-trivial geometries, the 1D emitters were therefore equally distributed azimuthally with respect to the symmetry axis, conserving the radial and vertical position and velocity components. The higher the number of 1D emitters, the better the sampling of the 3D emitter. It is enough to equally distribute only the emitting cells to calculate Doppler boosting (but not maps), which somewhat reduces the computational costs. This transformation only affects the radiative part of the code; the particle energy distributions in each cell or streamline are unaffected.

\subsubsection{Emission}\label{emission}

The luminosity per frequency unit (spectral luminosity) and per solid angle of the synchrotron emission in the FF for each cell and in the optically thin case was computed following \citep{bookPacholczyk}
\begin{equation}
L'_{\Omega'}(\epsilon') = c_3B'\sin{\theta'}\int_{E'_{\rm min}}^{E'_{\rm max}} N(E^*)F(x^*)\mathrm dE^*{\rm ,}
\end{equation}
where $N(E^*)$ is the cell particle energy distribution, $E^*$ the particle energy, $x^*=\epsilon'/\epsilon^*_c$, $\epsilon^*_c = c_1 h B'\sin{\theta'}E^{*2}$ the critical energy, $c_1 = {3e}/{4\pi m_e^3c^5}$, $c_3 = {\sqrt{3} e^3}/{4\pi h m_ec^2}$, $h$ the Planck constant, and $\theta'$ the angle between $\overrightarrow{B'}$ and the direction towards the observer.
The function $F(x)$ is defined through an integral of the $K_{5/3}(z)$ Bessel function, but we adopted the following approximation \citep[e.g.][]{bookAharonian}, which is valid in the interval $0.1 < x < 10$:
\begin{equation}
F(x) = \int_x^\infty K_{5/3}(z) \textrm dz \simeq 1.85\,x^{1/3} \exp(-x)\,.
\end{equation}
The orientation of $\overrightarrow{B'}$ is not fully defined in our approach, and for simplicity we made the approximation $B'\sin{\theta'}\rightarrow B'\sqrt{2/3}$. 
As particles move isotropically in the FF, the SED in that frame can be computed simply through 
\begin{equation}
\epsilon'L'(\epsilon')=4\pi \epsilon'L'_{\Omega'}(\epsilon')\,,
\end{equation}
where $L'(\epsilon')$ is the spectral luminosity of each cell in the FF.

The IC radiation was computed in the FF using the kernel for the emission rate of electrons of energy $E'$ interacting
with monodirectional target photons of energy $\epsilon'_0$ with an angle $\theta'$ \citep[e.g.][]{AhaAto81,bookAharonian,kak14}:
\begin{equation}
\frac{\mathrm d\bar{n}'(\theta',\epsilon')}{\mathrm d\epsilon'\mathrm d\Omega'} =  \frac{3\sigma_T}{16\pi \epsilon_0'E'^2}\,
\left[1+\frac{z^2}{2(1-z)}-\frac{2z}{b_{\theta'}(1-z)}+\frac{2z^2}{b_{\theta'}^2(1-z)^2}\right]\,{\rm ,}
\label{kernel}
\end{equation}
where $b_{\theta'} = 2(1-\cos\theta')\epsilon_0'E'$, and $z = \epsilon'/E'$. Convolving Eq.~(\ref{kernel}) with $N'(E')$ and the target photon energy distribution density in the FF, $n'_0(\epsilon'_0)$, the IC SED can be obtained:
\begin{equation}
\epsilon' L(\epsilon') = 4\pi c\epsilon'^2 \int_{E_{min}'}^{E_{max}'}\mathrm dE'\int_{E'_{\rm min}}^{E'_{\rm max}} \mathrm d\epsilon'_0 \frac{\mathrm d\bar{n}'(\theta',\epsilon')}{\mathrm d\epsilon'\mathrm d\Omega'} N'(E') n'_0(\epsilon'_0)\,{\rm .}
\end{equation}
A black body, centred at the star location, $(0,0.3\times 10^{15}$)~cm, was adopted to obtain $n_0(\epsilon_0)$, which in this case $n_0(\epsilon_0)=n'_0(\epsilon'_0)$, with a temperature and luminosity of $T_*=3\times 10^3$~K and $L_*=3\times 10^{36}$~erg~s$^{-1}$, respectively, typical for a red giant. Other photon fields such as the cosmic microwave background, the radiation from an accretion disk, the starlight from the galaxy, or radiation from the jet itself, were not taken into account because we focus here on the dominant photon field on the spatial scales of the calculations: the stellar photon field. If the non-thermal particles were followed up to cover larger spatial scales, however, other photon fields might be dominant and therefore should included in the calculations (see Sect.~\ref{discu}).

The derived SEDs were obtained for each cell in the FF. Owing to relativistic and light retardation effects, the photon energy and SED as seen by the observer have to be transformed as $\epsilon=\delta\epsilon'$ and $\epsilon L(\epsilon)=\delta^4 \epsilon^\prime L^\prime(\epsilon^\prime)$, where $\delta=1/\Gamma(1-\beta\cos{\theta_{\rm obs}})$, with $\theta_{\rm obs}$ being the angle between the flow and the observer direction in the laboratory frame (LF). For the IC, the angle $\theta'$ is also to be transformed from the LF. Further details on the transformation of the relevant angles can be found in Appendix~\ref{traang}.
For a dense, external target photon field, the gamma rays produced might be absorbed through pair creation (functionality included in the code). Electromagnetic cascading in the emitter environment may be also relevant in some situations. In the present scenario opacities are well below 1 in the considered cases, therefore
gamma-ray absorption is negligible.

\subsection{Results}

We applied the radiative code described in Sect.~\ref{sec:radiative} in conjunction with the RHD results of Sect.~\ref{sect:hydro} to compute the radiation in the two different scenarios: the star-jet interaction in the steady and the perturbed states. We let three parameters vary: the angle between the jet axis and the line of sight, $\phi$; the initial ratio of Poynting-to-matter energy flux (roughly the ratio of magnetic-to-thermal pressure just downstream of the jet shock), $\chi_B$; and the height in the jet, with respect to the jet base, where the jet-star interaction takes place: $z_{\rm int}$.The values considered for these parameters are summarized in Table \ref{tab:freepara}. We recall that the simulations were computed with a purely RHD code, meaning that the magnetic field is not dynamically relevant and only high enough to allow the plasma to behave as a fluid. However, we adopted  $\chi_B=10^{-1}$ in some cases, which is still small enough to avoid violating the low-$B$ condition, to have an example of a relatively high $B$-case.

\begin{table}
\centering
\begin{tabular}{c c}
\hline\hline
Parameter & Set of values\\
\hline
Jet observation angle $\phi$    & 0, 45, 90, $135^\circ$ \\
Fraction $\chi_B$                       & $10^{-4}, 10^{-1}$\\
Height $z_{\rm int}$            & 1, 10, 100, 1000 pc \\
\hline
\end{tabular}
\caption{Set of parameters for the two scenarios considered.}
\label{tab:freepara}
\end{table}

\subsubsection{Scalability of the results}

Throughout the paper, the jet power, $L_0$, opening angle \citep[taken $\sim 1/\Gamma_0$, as in][see references therein]{Bos15}, and wind thrust, $\dot{M}v_{\rm sw}$, were fixed to the following reference values: $\approx 10^{44}$~erg~s$^{-1}$, $0.1$~rad, and $1.3\times 10^{25}$~dyn, respectively. The star temperature and luminosity were also fixed to those of a red giant (see Sect.~\ref{emission}). However, the obtained results for the different values of $\phi$, $\chi_B$, and $z_{\rm int}$ explored can be easily generalized if $\Gamma_0$ and $T_*$ are fixed, and $L_*$ is approximated as $\sim \dot{M}v_{\rm sw}c$ \citep{Bos15}. An additional assumption to perform the generalization is that the radiation on the particle energy distribution (e.g. through synchrotron 
self-Compton) or the radiation itself (e.g. through internal pair creation) do not significantly interact. Under these conditions, the SEDs and mapped quantities presented below can be scaled as follows:

(i) The spectrum does not change if the escape-to-radiative timescale ratio, $$t_{\rm esc}/t_{\rm rad}\propto \sqrt{\dot{M}v_{\rm sw}L_0}/z_{\rm int}\,{\rm ,}$$ is constant, where $t_{\rm rad}\propto R_{\rm CD}^2/\dot{M}v_{\rm sw}$ and $t_{\rm esc}\propto R_{\rm CD}$. The quantity $t_{\rm rad}$ typically depends on the particle energy, which means that the timescale ratio has to be computed by fixing this energy to some particular value. The quantity $t_{\rm esc}$ would correspond to the typical timescale required for particles to escape the emitting region \citep[see][]{Bos15}.

(ii) The SED normalization changes as \citep{Bos15}$$\propto (R_{\rm CD}/R_{\rm j})^2L_0\propto \dot{M}v_{\rm sw}\,,$$ 
where $R_{\rm j}$ is the jet radius.
In the adiabatic regime of the emitter, the SED normalization is also $\propto t_{\rm esc}/t_{\rm rad}$. 

(iii) Thus, in the adiabatic regime, the non-thermal luminosity, either synchrotron or IC, can be simply scaled as $$\propto (\dot{M}v_{\rm sw})^{3/2}L_0^{1/2}/z_{\rm int}\,{\rm ,}$$ which is the product of the dependences stated in (ii). 

These relations are more refined versions than, but based on, those presented in Eqs.~(6) and (9) in \cite{Bos15}.

For the jet Lorentz factor $\Gamma_0$, \cite{Bos15} indicated that the normalization of the observer luminosity should approximately scale as $\propto \Gamma_0^2$. However, this is strictly valid in the ultra-relativistic regime, and far from the shock and/or under fast expansion of the streamlines; the scaling is less sensitive
to $\Gamma_0$ for relatively slow expansion and in the simulated region. 

\subsubsection{Spectral energy distributions and radiation maps}

\begin{figure} 
        \centering
        \includegraphics[width = \hsize]{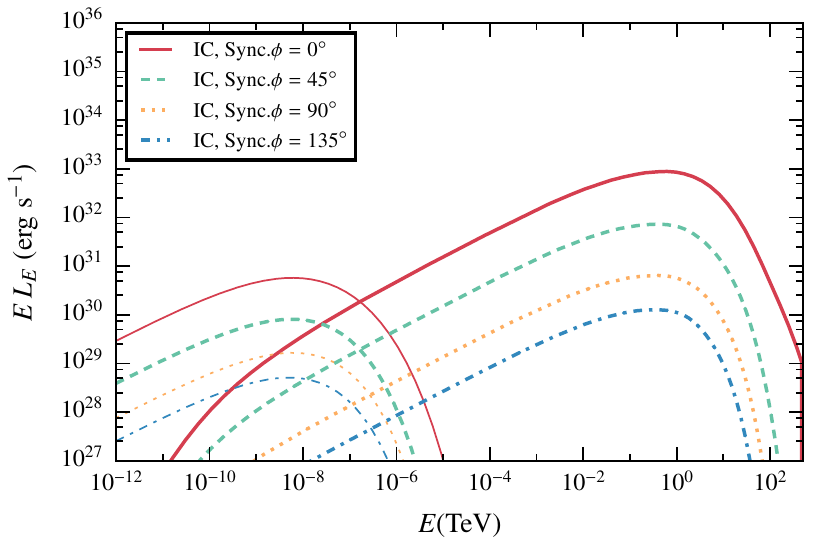}
        \caption{Synchrotron (thin) and IC (thick line) SEDs for the jet-star interaction in steady-state, taking $z_{\rm int}= 10$~pc, $\chi_B = 10^{-4}$, and for $\phi=0^\circ$, $45^\circ$, $90^\circ$ and $135^\circ$.}
    \label{fig:lowB}
   \end{figure}

\begin{figure} 
        \centering
        \includegraphics[width = \hsize]{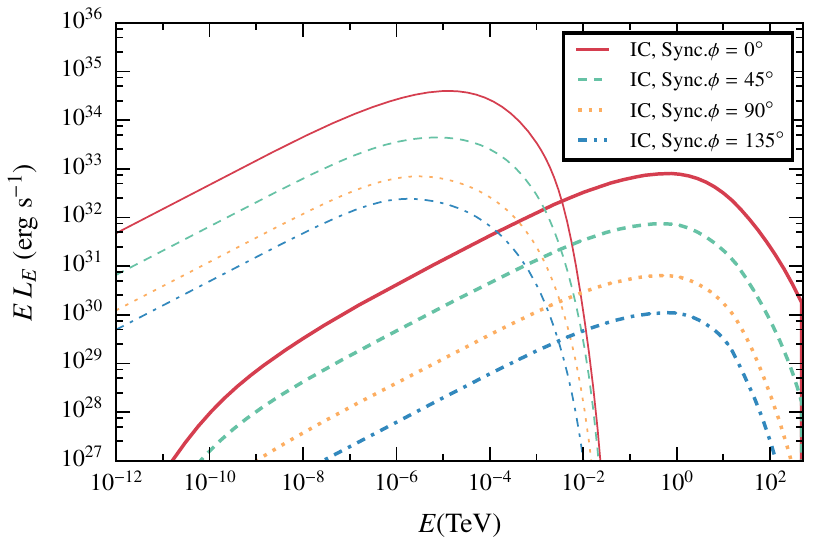}
        \caption{Same as in Fig.~\ref{fig:lowB}, but for $\chi_B=10^{-1}$.}
    \label{fig:highB}
   \end{figure}

Figures \ref{fig:lowB} and \ref{fig:highB} show the observer synchrotron and IC SEDs for the star-jet interaction in the steady-state, for $\chi_B = 10^{-4}$ and $10^{-1}$, $\phi=0^\circ-135^\circ$, and $z_{\rm int}=10$~pc. For high magnetization, synchrotron emission strongly dominates and reaches much higher photon energies. However, even in this case, most of the particle energy distribution is dominated by advection escape, and therefore the synchrotron and IC SED shapes do not depend significantly on $\chi_B$. The advection escape dominance also implies that non-thermal particles behave as an adiabatic flow and that the ratio of non-thermal-to-thermal energy will approximately keep constant along the streamlines.
As seen in the figures, the emission is significantly boosted for the jet on axis. Interestingly, for $\chi_B=10^{-1}$ the synchrotron SEDs for high $\phi$-values present softer spectra
because particles coming from the more Doppler-boosted outer lines (thus beamed away for high $\phi$) have higher maximum photon energies. This effect is not as clearly seen in the IC SED because the softening of the SED is already strong as a result
of the Klein-Nishina (KN) effect in the cross section.

\begin{figure} 
        \centering
        \includegraphics[width = \hsize]{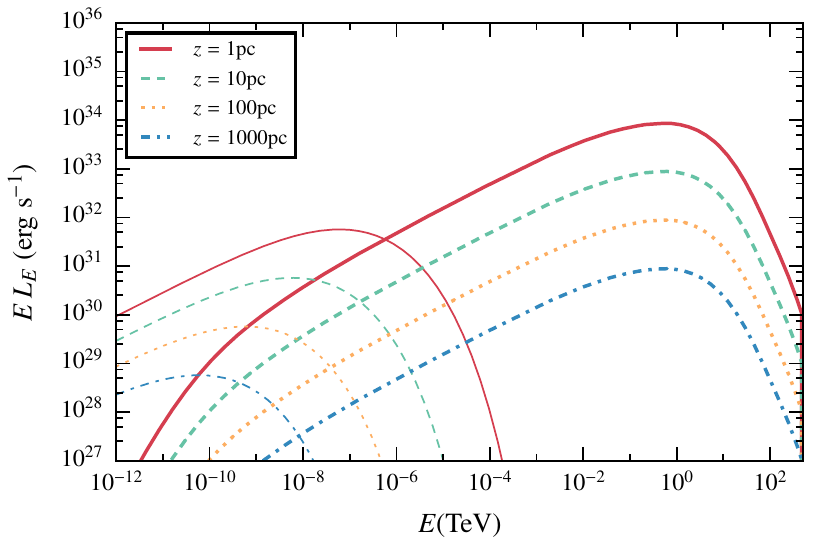}
        \caption{Synchrotron (thin) and IC (thick line) SEDs for the jet-star interaction in steady-state, taking $\phi=0^\circ$, $\chi_B = 10^{-4}$, and for
$z_{\rm int}=1$, 10, 100 and 1000~pc.}
    \label{fig:plotz}
   \end{figure}

In Fig.~\ref{fig:plotz} the observer synchrotron and the IC SED are shown for different values of $z_{\rm int}$ from 1 to 1000~pc, and for $\phi=0^\circ$, $\chi_B=10^{-4}$, and $\phi=0^\circ$. It is clear from the figure that the closer to the jet base, the higher the ratio of radiative versus advection escape, which as mentioned is $\propto 1/z_{\rm int}$.

\begin{figure*} 
        \centering
                \includegraphics{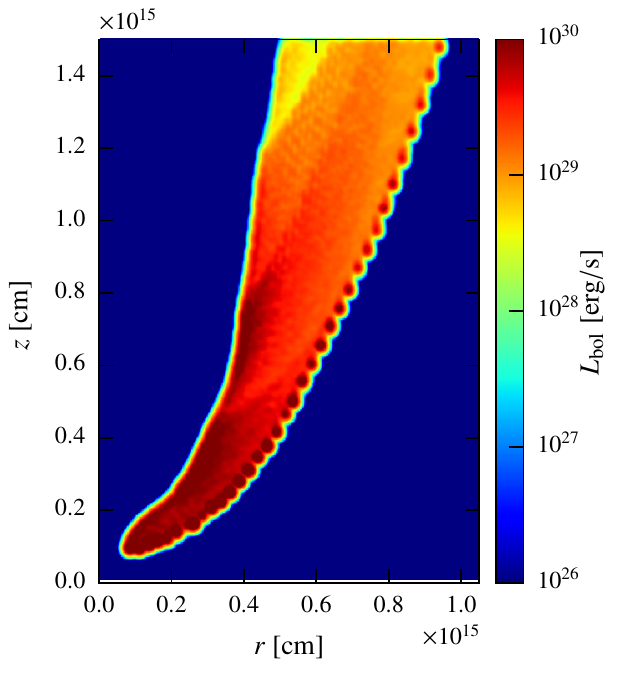}\hspace{0.15\textwidth}\includegraphics{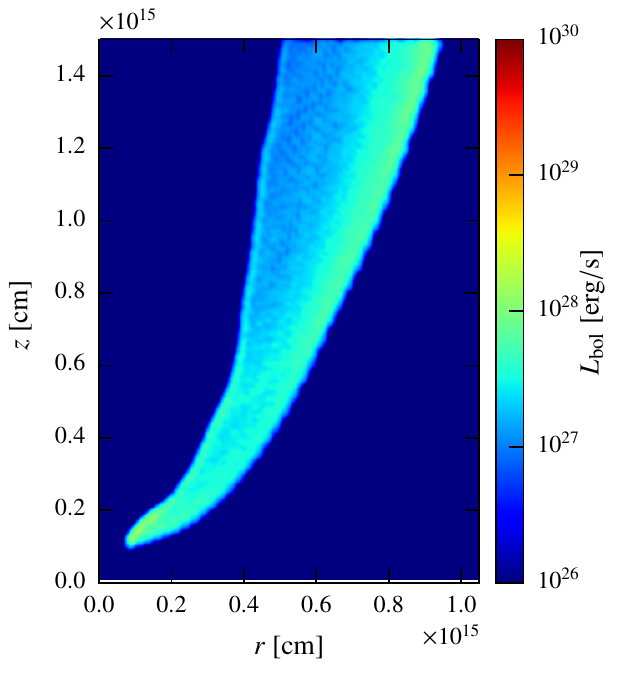}
        \caption{Map in the $rz$-plane of the distribution of IC (left), and synchrotron (right), bolometric luminosity per cell, for the jet-star interaction in steady-state. The adopted parameters are $\phi = 0$, $\chi_B=10^{-4}$ and $z = 10$~pc.}
    \label{fig:maps}
   \end{figure*}

Figure~\ref{fig:maps} presents the distribution of the bolometric luminosity per cell in the $rz$-plane for both synchrotron and IC, taking $\chi_B=10^{-4}$, $z_{\rm int}=10$~pc, and $\phi=0^\circ$. We note that because of the azimuthal geometry, the computational cells correspond to annular physical regions. The maps show that the synchrotron emission is more widely distributed than IC emission. This is because the IC target photon field is concentrated towards the star. 

\begin{figure} 
        \centering
        \includegraphics[width = \hsize]{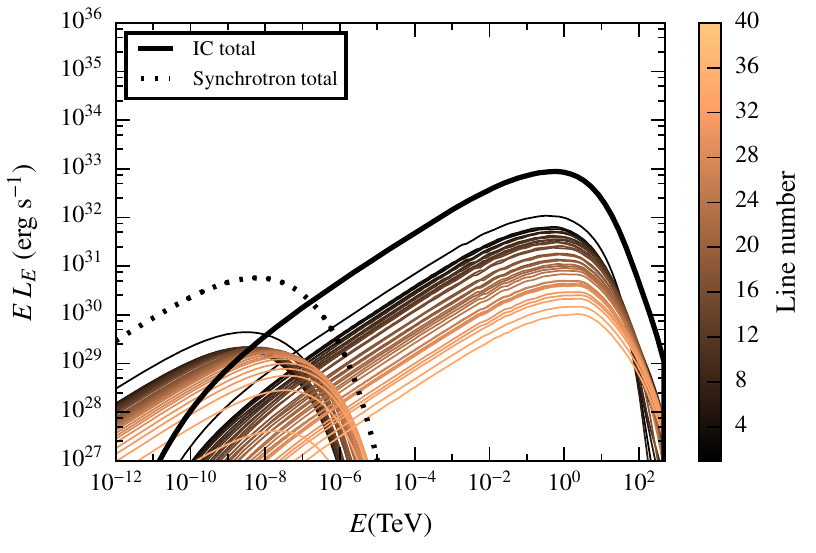}
        \caption{Synchrotron (dotted) and IC (solid line) SEDs for the different streamlines (thin grey lines), and the sum of all of them (thick black line) for the jet-star interaction in steady-state and $z_{\rm int}=10$~pc, $\chi_B=10^{-4}$ and $\phi=0^\circ$.}
    \label{fig:dist1}
   \end{figure}

\begin{figure} 
        \centering
        \includegraphics[width = \hsize]{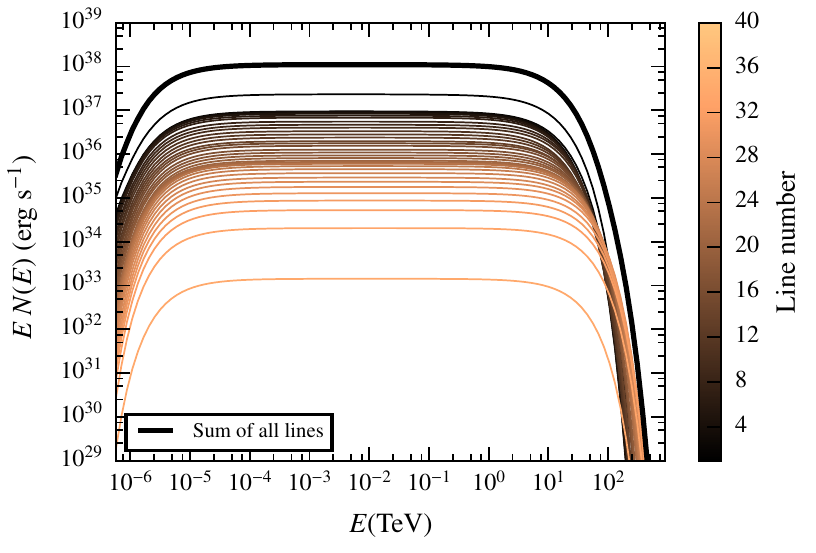}
        \caption{Particle energy distributions for the different streamlines (thin grey lines), and the sum of all of them (thick black line) for the jet-star interaction in steady-state and $z_{\rm int}=10$~pc, $\chi_B=10^{-4}$.}
    \label{fig:dist2}
   \end{figure}

Finally, in Figs.~\ref{fig:dist1} and \ref{fig:dist2} we show the contribution of the different lines for $z_{\rm int}=10$~pc, $\chi_B=10^{-4}$, and $\phi=0^\circ$, to the observer synchrotron and IC SED, and total energy distribution in the LF, respectively. The contribution to the emission varies substantially between different streamlines, depending on the non-thermal particle content, the role of adiabatic cooling or heating along the lines, the flow velocity and direction, the local magnetic field, and the relative position with respect to the source of target photons.

\begin{figure} 
        \centering
        \includegraphics[width = \hsize]{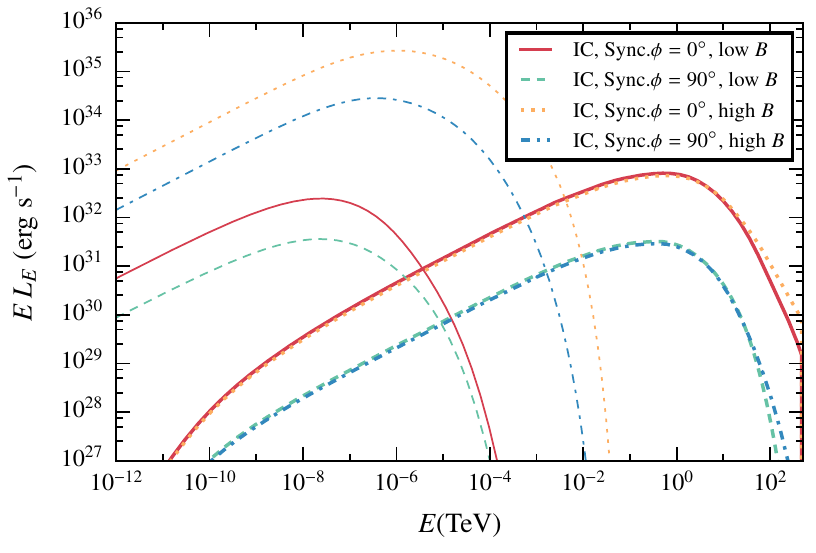}
        \caption{Synchrotron (thin) and IC (thick line) SEDs for the jet-star interaction in the perturbed state, taking $z_{\rm int}= 10$~pc, $\chi_B = 10^{-4}-1$, and $\phi=0^\circ-135^\circ$.}
    \label{fig:plotbrac}
   \end{figure}

\begin{figure*} 
        \centering
                \includegraphics{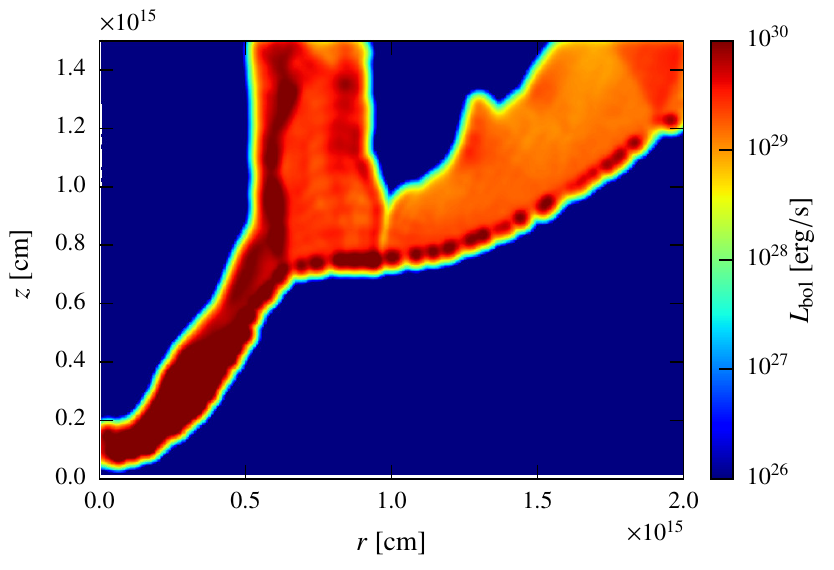}\includegraphics{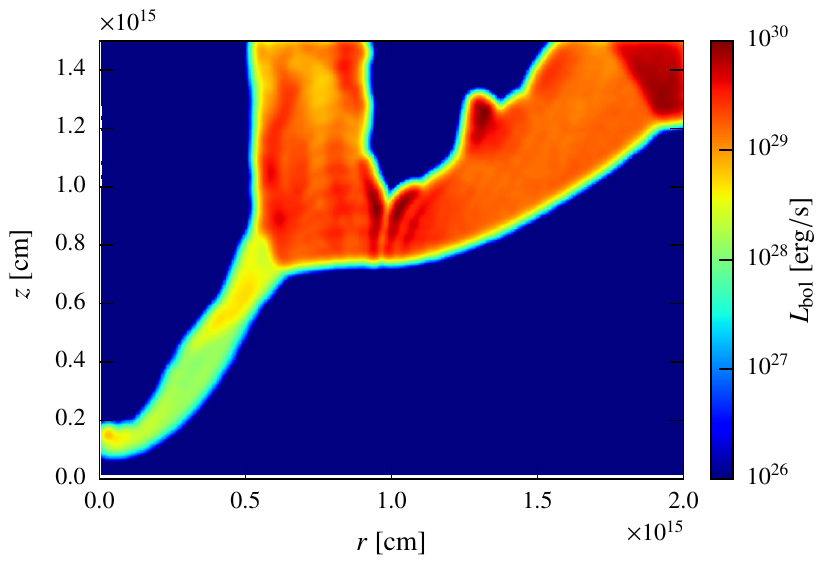}
        \caption{Map in the $rz$-plane of the distribution of IC (left), and synchrotron (right), bolometric luminosity per cell for the jet-star interaction in the perturbed state. The adopted parameters are $\phi = 0$, $\chi_B=10^{-4}$ and $z = 10$~pc.}
    \label{fig:mapsbrac}
   \end{figure*}

Figure \ref{fig:plotbrac} shows the observer synchrotron and IC SEDs for the star-jet interaction in the perturbed state, for $\chi_B = 10^{-4}$ and $10^{-1}$, $\phi=0^\circ-135^\circ$, and $z_{\rm int}=10$~pc. Figure~\ref{fig:mapsbrac} presents the distribution in the $rz$-plane of the bolometric luminosity per cell for both synchrotron and IC, taking $\chi_B=10^{-4}$, $z_{\rm int}=10$~pc, and $\phi=0^\circ$. 

\begin{figure*} 
        \centering
        \includegraphics{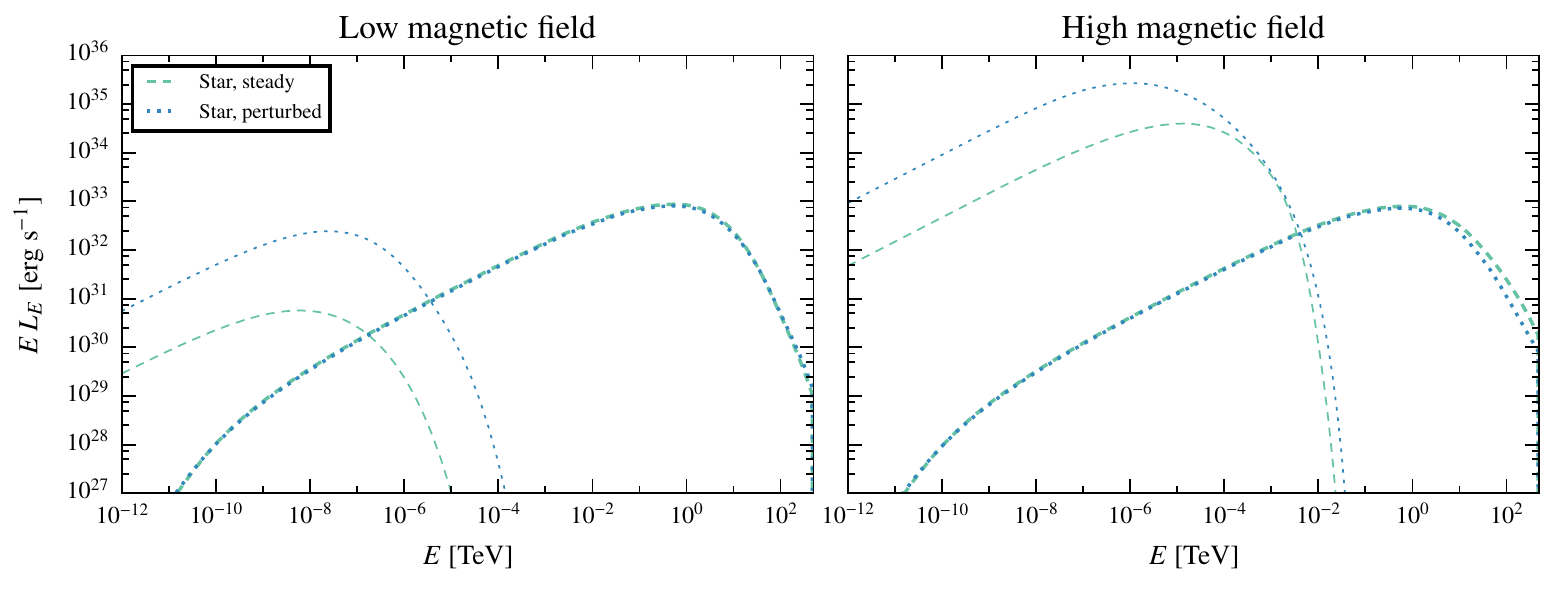}
        \caption{Synchrotron (thin) and IC (thick line) SEDs for the star-jet steady and perturbed state, taking $\phi=0^\circ$ and $z_{\rm int}=10$~pc, for $\chi_B=10^{-4}$ (left) and $10^{-1}$ (right).}
    \label{fig:comparison}
   \end{figure*}

Figure~\ref{fig:comparison} shows the synchrotron and IC SEDs for the two cases studied in this work, in the low- and high-magnetization case and for $\phi=0^\circ$ and $z_{\rm int}=10$~pc. The figure
shows that the synchrotron emission is significantly higher for the star-jet interaction in the perturbed state because a larger section of the jet is affected by the stellar wind, which is apparent from comparing Figs.~\ref{star_density_lines} and \ref{star_pert_density_lines} in Sect.~\ref{sect:hydro}. This implies that more jet energy available for radiation. On the other hand, the IC radiation levels change very little when compared to the synchrotron levels because the target photon density significantly drops with distance from the star. In the high-magnetization case, the differences between the two studied cases are smaller than for a low magnetization, and the synchrotron SED of the perturbed case shows a softer spectrum at the highest energies. This is most likely related to the high magnetic field, which through severe synchrotron cooling prevents the most energetic electrons from reaching regions of higher Doppler-boosting. Otherwise, the radiation of these particles would have led to more flux at the higher synchrotron energies and thus to a harder spectrum.

\section{Conclusions}\label{conc}

The results of the radiation calculations carried out using hydrodynamical information are in the line of those presented in \cite{Bos15}. In that work, it was noted that the effective size of the obstacle is much larger than the distance at which this and the jet flow collided (i). It was also pointed out that Doppler-boosting plays a significant role (ii). All this is confirmed here: 

(i) For instance, for the jet-star interaction in steady-state with $\phi=0^\circ$, $z_{\rm int}=10$~pc, and $\chi_B=10^{-4}$, the total observer luminosity is $\approx 3.5\times 10^{31}$ and $5\times 10^{33}$~erg~s$^{-1}$ for the synchrotron and IC components, respectively, whereas the synchrotron and IC luminosities from a region $r<R_{\rm CD}$ are $\approx 2\times 10^{29}$ and $8\times 10^{31}$~erg~s$^{-1}$, respectively, which is a factor $\sim 100$ smaller than for the whole grid. 

(ii) In addition, the total synchrotron and IC luminosity in the FF are $\approx 1.3\times 10^{30}$ and $2\times 10^{32}$~erg~s$^{-1}$, respectively, which is a factor $\sim 30$ smaller than the observer values when Doppler-boosting is not accounted for. It is worth noting that jets on kpc and larger scales may be less relativistic, as considered in \cite{BedBan15}, but closer to the galaxy centre, they are likely to have much higher Lorentz factors. Farther out, jet-boundary instabilities, shear-layer development, recollimation shocks, and mass-loss from stellar winds work together to decelerate the jets \citep[see e.g.][]{Per14,Per15}.

As expected from previous work \citep[e.g.][]{abr13,BedBan15,Bos15}, we find that advection escape dominates radiation losses in the star-jet interactions studied, that is, for moderately powerful jets and stellar winds. Radiation peaks from X-rays to MeV energies depending on $\chi_B$ for synchrotron emission, and in the 100-1000~GeV range for IC, with the cooler, either older or less massive, stars yielding the higher SED peak energy ($\sim 1$~TeV, as in the present work). A study of collective interactions of many stars with the jet \citep[see][for massive stars]{abr13} is under way to also account for other stellar populations such as evolved stars and for the increased effective section of the obstacles and relativistic effects. However, \cite{Bos15} predicted significant collective emission, for which they took  the stellar populations of the inner regions of the radio galaxy M~87 as a reference. These predictions receive further support here because we obtain similar IC fluxes and general behaviour for individual interactions, although we note that a more detailed prescription of the stars and their winds in the inner regions of AGN is required. 

 Instabilities might play an important role, and the shocked flow structure might be relatively unstable. This may enhance the emission for some individual jet-star interactions, although the collective emission would likely be at about steady-state radiation levels. Otherwise, for interactions taking place in the innermost jet regions and bright enough to be detectable by themselves, instability growth could temporary increase the emission levels and induce variability on scales $\gtrsim R_{\rm CD}/v_{\rm sw}$ in addition to other types of  variability, such as that associated with jet crossing. It is worth noting that a longitudinal magnetic field in the jet would reduce the instability
growth.

\section{Final remarks}\label{discu}

The question whether electrons or $e^\pm$-pairs are indeed accelerated in the jet shock is a key point. \cite{Bos15} showed that for  luminosities per interaction similar to those obtained by us and under the same interaction conditions, collective jet-star interactions could yield detectable levels of emission in M~87, for instance. Therefore, our results together with those from \cite{Bos15} indicate that for acceleration efficiencies of $\chi_{\rm NT}\gtrsim 0.1$, collective star-jet interactions may be detectable in AGN in gamma rays unless (i) an AGN jet is a stronger emitter through a different mechanism, and/or (ii) the source is far and the jet seen off-axis, and/or (iii) the jet power is low. It is worth noting that the need of a relatively high $\chi_{\rm NT}$-value applies particularly to the low-$B$ scenario because for higher magnetizations the synchrotron component could overcome the IC in the 100~MeV region, which would weaken this requirement. 
   
The limited size of the grid makes accounting for all the emission produced in the interaction region difficult, in particular for the synchrotron radiation. \cite{Bos15} noted that non-negligible kinetic to internal energy transformation takes place far from the simulation axis, although with a weak $r$-dependence. However, Doppler-boosting effects are expected to be strong far from the obstacle, and synchrotron emission can come from farther regions than IC because the latter is affected by stellar field dilution with distance from the star. Larger grid simulations or jet-scale semi-analytic calculations are required to determine the emission contribution of these farther regions. As a reference, we point out that for the case with $\phi=0^\circ$, $z_{\rm int}=10$~pc, and $\chi_B=10^{-4}$, $\sim 10$\% of the synchrotron and 30\% of the IC total luminosity come from a distance $3\times R_{\rm CD}$ from the simulation axis. Therefore, the trend of the IC emission indicates convergence, but the synchrotron emission may be still far from that. This is linked to the fact that our simulations yield very low adiabatic cooling rates. The dominant source by far of non-thermal energy loss is advection escape from the grid. This is consistent with the modest density decrease with $z$ in the shocked jet flow, as the density maps in Sect.~\ref{results} illustrate. Our results are complementary to those in \cite{BedBan15}, where jet emission from particles accelerated in jet-obstacle interactions was computed by accounting for the jet $B$ and photon fields that are relevant on larger spatial scales, such as the galaxy bulge or the CMB. A larger grid would also help to study more unstable configurations, such as adopting a higher density wind-jet contrast or increasing the resolution for a more realistic setup, although the computation time required would severely increase. Finally, 3D simulations that account for the star motion also need to be carried out.
   
\begin{acknowledgements}

We acknowledge support by the Spanish Ministerio de Econom\'ia y Competitividad (MINECO) under grants
AYA2013-47447-C3-1-P, and MDM-2014-0369 of ICCUB (Unidad de Excelencia 'Mar\'ia de Maeztu').      
This research has been supported by the Marie Curie Career Integration Grant 321520.
V.B-R. also acknowledges financial support from MINECO and European Social Funds through a Ram\'on y Cajal fellowship.
X.P.-F. also acknowledges financial support from Universitat de Barcelona and Generalitat de Catalunya under grants APIF and FI (2015FI\_B1 00153), respectively.
D.K. acknowledges financial support by a grant-in-aid for Scientific Research (KAKENHI, No. 24105007-1)  from the Ministry of Education, Culture, Sports, Science and Technology of Japan (MEXT).
M.P. is a member of the working team of projects AYA2013-40979-P and AYA2013-48226-C3-2-P, funded by MINECO.
\end{acknowledgements}

\bibliographystyle{aa}
\bibliography{ALLreferences}

\begin{appendix}

\section{Streamline calculations}\label{strlin}

The simulations are axisymmetric, and by assuming that the flow is stationary, which is approximately valid for most of the jet material within the grid, we can therefore introduce a {\it \textup{stream function}}, $\Phi$, as
\begin{equation}
\Gamma\rho \mathbf{v}={\nabla \Phi \times {\mathbf e}_\psi\over 2\pi r}\,.
\end{equation}
The entire flow can be divided into a set of regions, $V_i$, based on the value of the stream function:
$\Phi_{i-1}<\Phi(r,z)\le\Phi_{i}$, where $1\le i\le N$ and $\Phi_0=0$. Physically, each of these regions fulfils the condition that the same amount of matter flows through their cross section. Since at the bottom boundary of the computational domain the quantities $v_r$ and $\Gamma\rho v_z$ are constant, with $v_r=0$, the stream function can be computed on this surface from $d \Phi =2\pi r \Gamma \rho\left( v_z dr - v_r dz \right)= \Gamma_{\rm 0}\rho_{\rm 0} v_{\rm 0} dS_0$, where $dS_0=2\pi rdr$ is a surface element. This means that at the bottom of the computational domain the cross sections of the regions $V_i$ correspond to a sequence of annular rings: $r_{i-1}<r\le r_{i}$, with $r_0=0$, $r_{N}=l_r$, and $\Phi_i=\pi r_i^2 \rho_{\rm 0} v_{\rm 0}\Gamma_{\rm 0}$. 

These regions $V_i$ form layers with axial symmetry in the 3D space. If their thickness is small enough, they can be considered homogeneous in the directions perpendicular to the fluid motion, and the properties of the non-thermal particles in each layer will depend only on the time spent by the fluid element of interest since its injection into the computational domain. In particular, the number density and energy distribution of the non-thermal particles responsible for the synchrotron and IC emission of the jet in a region $V_i$ depend only on this time. To compute the energy distribution numerically, we selected a streamline for each volume element $V_i$ and solved the particle transport equation along this line using the momentarily snapshot of the plasma properties obtained with our hydrodynamic simulations (for details see Appendix~\ref{injpart}).

Each streamline was divided in segments by an equidistant time step $\Delta t$:
\begin{align}
& r_{{\rm s},j} = r_{{\rm s},j-1}+v_{r,j-1}\Delta t\rm{,}\nonumber\\
& z_{{\rm s},j} = z_{{\rm s},j-1}+v_{z,j-1}\Delta t\rm{,}
\end{align}
where $\Delta t$ is the smallest cell size divided by the highest speed in the simulation. The initial point for the streamline in the region $V_i$ was selected as $r_{{\rm s},0}=(r_{i-1}+r_{i})/2$ and $z_{{\rm s},0}=0$. This division of the streamlines leads to the splitting of the regions $V_i$ into annular cells. The cross sections transversal to the flow velocity of the cells can be computed assuming conservation of the energy crossing those sections per time unit,
\begin{equation}
S = S_0\frac{\rho_0 \Gamma_0^2 h_0 v_0}{\rho \Gamma^2 h v}\rm{,}
\end{equation}
where $h$ and $v$ are the specific enthalpy and the modulus of the three-velocity, respectively. The subscript $0$ denotes the conditions at the bottom boundary of the region $V_i$, that is, in particular,

\begin{equation}
S_0 = \pi\left(r_{i}^2-r_{i-1}^2\right)\,\rm{.}
\end{equation}

After the next point along the streamline, $(r_{{\rm s},j},z_{{\rm s},j})$, was obtained, the hydrodynamic parameters ($v_{r,j}$, $v_{z,j}$, $\rho_{j}$, and $p_{j}$) were computed through a bilinear interpolation of the corresponding values from the neighbour cells \citep{bookPressTeukolskyVetterling}. This was done iteratively until the fluid element followed along the streamline left the simulation grid.

The magnetic field at the inlet of the computational domain in the FF, $B'_0$, was calculated assuming that the Poynting flux is equal to a fraction $\chi_B$ of the matter energy flux, which yields
\begin{equation}
\frac{B_0^{'2}}{4\pi} = \chi_B  \rho_0 h c v\,.
\label{eq:B_evolution}
\end{equation}
Then, from Eq.~(\ref{eq:B_evolution}), the magnetic field at each cell of the streamline can be derived assuming that it is frozen into the plasma and is perpendicular to its direction of motion:
\begin{equation}
B' = B'_0\sqrt{\frac{\rho v_0 \Gamma_0}{\rho_0 v \Gamma}}\,.
\end{equation}

Finally a downsampling of each streamline through interpolation was performed to reduce the computation time of the radiative code, resulting in streamlines sampled with up to $200$ equidistant cells. 

\section{Non-thermal particles}\label{injpart}

\subsection{Injection of particles}

 Shocks are assumed here to be the sites of particle acceleration. Using streamline data, we can compute the energy rate of accelerated particles (or non-thermal luminosity) injected in cells were shocks occur. Injection is thus assumed to take place when there is an increment in the internal energy and a decrement in the flow velocity between two consecutive cells to approximately capture shocks. The rate of energy injection in the form of accelerated particles in the cell volume in the FF, $L'_{\rm NT}$, is taken to be a fraction $\chi_\text{NT}\le 1$ of the generated internal energy per second in the cell $k$ ($\dot{U'_k}$), that is, the time derivative of the ${00}$-component of the energy-momentum tensor times volume in the FF:
\begin{equation}
L'_{\rm NT}=\chi_{\rm NT}\dot{U'}_k=\chi_{\rm NT}v_k\Gamma_k^2
\label{eqnt}
\end{equation}
$$
\times\,\left(\left(S_+\Gamma_+^2h_+\rho_+c^2\left(1+\frac{v_+^2}{c^2}\right)\right)-
S_-\Gamma_-^2h_-\rho_-c^2\left(1+\frac{v_-^2}{c^2}\right)\right)\,{\rm ,}
$$
where $v_{+/-}$ and $\Gamma_{+/-}$ are the velocities and Lorentz factors in the cell right and left boundaries, respectively. Equation~(\ref{eqnt}) is valid for a steady-state fluid. 

The normalization of the source function for non-thermal particles, $Q'$, should satisfy the following condition:
\begin{equation}
\int E'Q'(E')dE'=L'_{\rm NT}\,.
\end{equation}
At this stage, the expression we assume for $Q'$ is a power-law of index $-2$, typical for shock acceleration in different contexts (see below), with two cutoffs at high and low energies:
\begin{equation}
Q'(E') \propto E'^{-2} \left[\exp\left(\frac{-E'_{\rm c,low}}{E'}\right)\right]^5\exp\left(\frac{-E'}{E'_{\rm c,high}}\right)\,.
\end{equation}
The steep exponential cutoff towards low energies is adopted to avoid numerical artefacts at the lowest energies, and also to make the number of transrelativistic particles small. The low-energy cutoff energy is arbitrarily fixed here to $E'_{c,low}=1$~MeV. 

In addition to the fraction of energy going to non-thermal particles $\chi_{\rm NT}$, it is also required to set the timescale at which particles gain energy when accelerated to determine the highest particle energy, since the high-energy cutoff is fixed to the energy at which the acceleration timescale is equal to the shortest cooling or escape timescale. 
For the acceleration timescale, we assume that the process is linked to the jet shock and adopt a phenomenological prescription $t'_{\rm acc} = \eta E'/qB'c$ with $\eta=2\pi(c/v)^2$, which tends to $\sim 10$ when $v\rightarrow c$.
This is done for simplicity because particle acceleration in relativistic shocks is at present a complex and far from settled matter, and the simple prescription is enough for our purposes.
The prescription is loosely based on the acceleration timescale for non-relativistic strong shocks \citep[e.g.][]{Dru83}, although in our prescription $v$ refers to the flow velocity in the laboratory reference frame (LF), and not to the shock velocity in the reference frame of the medium upstream of the shock (as in particle acceleration theory). 
The cooling or escape timescales are given by the synchrotron timescale (dominant at the highest energies), $t'_{sync}=1/a_{\rm s}B'^2E'$ ($a_{\rm s}=1.6\times 10^ {-3}$, cgs units\footnote{Here we have averaged over the particle momentum-magnetic field pinch angle (see Sect.~\ref{emission}).}); and the Bohm diffusion timescale, $t'_{\rm diff} = R'^2_{\rm a}/2D'$, with $R'_{\rm a}$ being the typical size of the emitter, and $D'=cE'/3qB'$ being the Bohm diffusion coefficient. In this work, $R_{\rm a}=R'_{\rm a}/\Gamma$ is fixed to $3\,R_{\rm CD}$. The high-energy cutoff can be obtained by combining the different mentioned timescales:
\begin{equation}
E'_{\rm c,high} = \min\left(\frac{94}{\sqrt{B'\eta}}, \frac{5.6\times 10^{-10}B'R'_{\rm a}}{\sqrt{\eta}}\right)\,.
\label{eq:cutoff}
\end{equation}

When the internal energy increases, we set $\nabla(\Gamma\overrightarrow{v})$ to zero because in the injected luminosity we already take  the adiabatic heating that would take place into
account. Formally, non-thermal particle injection should take place only at shocks, and adiabatic cooling or heating would operate otherwise. We choose the adopted approach for simplicity,
however, as defining a shock accurately in streamlines is numerically more complex and demanding.

The emission results depend relatively weakly on the particle injection assumptions. This is true in particular for the IC luminosity in gamma rays, which is not significantly affected by $E_{\rm c,high}$ as long as it is $\gtrsim 100$~GeV, which is not a very demanding value. The relevant parameter is $\chi_{\rm NT}$, which is directly proportional to the normalization of the injected particle distribution and is not constrained.

\subsection{Particle energy distribution}

To compute the non-thermal particle evolution, we considered the hydrodynamical information constant over a time $\bar{t}$,  much shorter than the dynamical timescale of the simulations. During this time, we let the particles evolve until they reach a stationary solution for their energy distribution. This calculation is made in many short time steps:
\begin{equation}
\bar{t}=\sum\Delta \bar{t}_i=\sum\frac{1}{4}\min[\Delta x_{k,i}/v_{k,i}]\,{\rm ,} 
\end{equation}
where the subindex $i$ relates to the time step at time $\bar{t}_i$ in the LF, $k$ to the cell, and $\min[\Delta x_{k,i}/v_{k,i}]$ is the shortest cell crossing time at $\bar{t}_i$, with $\Delta x_{k,i}$ being the cell length. To compute the particle energy distribution at each (LF time) $\bar{t}_i$ and cell, the code, first order in space, time, and energy, follows three stages:

\begin{itemize}
\item First, the energy distribution after a $\Delta \bar{t}_i'=\Delta \bar{t}_i/\Gamma$ of the particles injected at $\bar{t}_i$ in the cell $k$, in the FF, is obtained from (dropping the indexes $i$ and $k$):
\begin{equation}
N'_1(E',\bar{t}) = \frac{1}{|\dot{E}|}\int_{E'}^{E'_{eff}} dE^* Q'(E^*)\,{\rm ,}
\label{eq:intQ}
\end{equation}
where $E'_{\rm eff}\le E'_{\rm c,high}$ is the energy that particles had before advancing the FF time a $\Delta \bar{t}'$, and is given implicitly by
\begin{equation}
\Delta \bar{t}' = \int_{E'}^{E'_{eff}} \frac{dE^*}{|\dot{E}'(E^*)|}\,{\rm .}
\label{eq:intt}
\end{equation}
Equation~(\ref{eq:intQ}) is the solution of the equation adapted from \cite{GinSyr64} to compute the evolution of particles in an homogeneous region,
\begin{equation}
\frac{\partial n'(E',\Delta \bar{t}')}{\partial \bar{t}'} + \frac{\partial (\dot{E}'(E')\, n(E',\Delta \bar{t}'))}{\partial E'} = Q'(E')\,, \label{eq:EDO dmitry}
\end{equation}
where $\dot{E}'$ is the particle energy-loss rate including all the relevant losses in the FF: IC \citep[e.g.][]{kak14}; synchrotron; and adiabatic cooling. The adiabatic losses are computed using the divergence of the spatial part of the flow four-velocity,
\begin{equation}
\dot{E}'_{\rm ad}(E') = -\frac{1}{3}\nabla(\Gamma\overrightarrow{v}) \cdot E'\,. 
\end{equation}

\item Second, even if no particles are accelerated in a cell, there are still particles that had come at $\bar{t}_{i-1}$ from the previous cell following an energy distribution $N'(E'_{\rm eff}, \bar{t}_{i-1})$. These particles evolve as
\begin{equation}
N'_2(E',\bar{t}_i) = N'(E'_{\rm eff}, \bar{t}_{i-1})\frac{\dot{E}'(E'_{\rm eff}, \bar{t}_{i-1})}{\dot{E}'(E', \bar{t}_i)}\,{\rm ;}
\label{eq:evolution}
\end{equation}
under steady cooling conditions, $\dot{E}'(E'_{\rm eff}, \bar{t}_{i-1})=\dot{E}'(E'_{\rm eff}, \bar{t}_{i})$.

\item After particles are evolved in energy, flow advection is taken into account to include the contribution from those evolved particles arriving from the previous cell, and remove the locally evolved particles that flow to the next one. 
In the cell $k$ the advection effect is
\begin{equation}
N'_3(E',\bar{t}_i)=N'_{k-1}(E',\bar{t}_i)\left(\frac{\Delta \bar{t} v_{k-1}(\bar{t}_i)}{\Delta x_{k-1}(\bar{t}_i)}\right)\\
-N'_{k}(E', \bar{t}_i)\left(\frac{\Delta \bar{t} v_{k}(\bar{t}_i)}{\Delta x_{k}(\bar{t}_i)}\right)\,{\rm ,}
\label{eq:advection}
\end{equation}
where 
\begin{equation}
N'_k(E',\bar{t}_i) = N'_1(E',\bar{t}_i) + N_2(E',\bar{t}_i)\,{\rm ,}
\end{equation}
and $(\Delta\bar{t} v_{k-1}(\bar{t}_i)/\Delta x_{k-1}(\bar{t}_i))$ and $(\Delta\bar{t} v_{k}(\bar{t}_i)/\Delta x_{k}(\bar{t}_i))$ are the fraction of particles of a given cell, $k-1$ and $k$, respectively, which left that cell.  
\end{itemize}

The computing tool described here can deal with non-linear processes, such as synchrotron self-Compton (SSC) because when the synchrotron radiation is computed, the particle distribution can be recalculated using a new ambient photon field that includes this component. Internal pair creation could be also taken into account as an additional source of injected particles, but this functionality has not been implemented yet. At present, nevertheless, these processes are not relevant in the
studied scenario: with respect to IC on locally produced radiation, as SSC, IC on the external photon field of the star dominates these other IC channels. On the other hand, pair creation through photon absorption in external or local photon fields has a very low probability and can be neglected.

\section{Transformation of angles to the fluid co-moving reference frame}\label{traang}

Several sources of target photons can play an important role in
inverse Compton scattering. However, given the small size of the
computational domain, we here only considered the
contribution from the star, which has been approximated as a
point-like emitter. Therefore, IC scattering proceeds in the
anisotropic regime \citep[see e.g.][]{BogAha00}, and we have to account for several effects that
involve relativistic transformations of angles. In particular,
particle cooling and scattering occur in the FF, which means
that to compute these
processes, we need to transform the photon field and scattering angle
to the FF.

We consider a 2D hydrodynamic flow and a coordinate system with
the $z$-axis coinciding with the symmetry axis. The source of target
photons is located at $\mathbf{r}_*=(0,0,z_*)$ and the observer is in
the $rz$-plane, that is, the line of sight is parallel to
\begin{equation}
\mathbf{n}_{\rm obs}=(\sin\alpha,0,\cos\alpha)\,.
\end{equation}
A fluid element
located at $\mathbf{r}=(r\cos\psi,r\sin\psi,z)$ (with $r$, $\psi$ and $z$ being
cylindrical coordinates) moves with  four-velocity
\begin{equation}
 \mathbf{u}=(\Gamma,u_r\cos\psi,u_r\sin\psi,u_z)=\Gamma(1,v_r/c\cos\psi,v_r/c\sin\psi,v_z/c)\,
\end{equation} 
along the direction represented by
\begin{equation}
\mathbf{n}_u={(u_r\cos\psi,u_r\sin\psi,u_z)\over\sqrt{u_r^2+u_z^2}}\,.
\end{equation}
Here, $\Gamma^2=1+u_r^2+u_z^2$ is the fluid element Lorenz factor.
Particles in the flow are illuminated by photons with
velocity directed along
\begin{equation}
\mathbf{n}_{\rm ph}={(r\cos\psi,r\sin\psi,z-z_*)\over
  \sqrt{r^2+(z-z_*)^2}}\,.
\end{equation}

For the non-thermal emission calculations, the following angle-related parameters are required: 
\begin{itemize}
\item[(i)] The Doppler factor is given by the following expression:
\begin{equation}
\delta={1\over \Gamma\left(1-\beta \mathbf{n}_u\mathbf{n}_{\rm obs}\right)}={1\over \Gamma-u_r\cos\psi\sin\alpha-u_z\cos\alpha}\,,
\end{equation}
where $\beta=\sqrt{u_r^2+u_z^2}/\Gamma$.
\item[(ii)] To compute the IC losses, we need to transform the photon field to the FF. This transformation is determined by the Doppler factor of the stellar photons \citep{kak14}:
\begin{equation}
{\cal D}_*={1\over \Gamma\left(1-\beta \mathbf{n}_u\mathbf{n}_{\rm ph}\right)}={1\over \Gamma-{u_rr+u_z(z-z_*)\over\sqrt{r^2+(z-z_*)^2}}}\,.
\end{equation}
\item[(iii)]  Gamma-gamma absorption on a field provided by a point-like source is determined by  the distance to the source of target photons and by the angle between  $\mathbf{n}_{\rm obs}$ and $\mathbf{n}_{\rm ph}$ \citep{kab08}:
\begin{equation}
\mathbf{n}_{\rm obs}\mathbf{n}_{\rm ph}={r\cos\psi\sin\alpha+(z-z_*)\cos\alpha\over \sqrt{r^2+(z-z_*)^2}}\,. 
\end{equation}
\item[(iv)] Finally, to compute anisotropic IC, the scattering angle in the FF is needed . To derive $\mathbf{n}_{\rm obs}'\mathbf{n}_{\rm ph}'$,
let $k=(k,k\mathbf{n}_{\rm obs})$ be the momentum of a photon propagating towards the observer, and $\omega=(\omega,\omega \mathbf{n}_{\rm ph})$ the momentum of a photon emitted by the star towards the fluid element. Since the scalar products $(ku)$ and $(\omega u)$ are Lorentz invariant and the four-velocity of the fluid element in the FF is $u'=(1,0)$, we obtain that
\begin{equation}
\omega'=\omega(\Gamma-\sqrt{u_r^2+u_z^2}\mathbf{n}_u\mathbf{n}_{\rm ph})=\omega{\cal D}_*^{-1}\,,
\end{equation}
and
\begin{equation}
k'=k(\Gamma-\sqrt{u_r^2+u_z^2}\mathbf{n}_u\mathbf{n}_{\rm obs})=k\delta^{-1}\,.
\end{equation}
Since $(k\omega)$ is an invariant,
\begin{equation}
k\omega(1-\mathbf{n}_{\rm obs}\mathbf{n}_{\rm ph})=k'\omega'(1-\mathbf{n}_{\rm obs}'\mathbf{n}_{\rm ph}')\,,
\end{equation}
and consequently,
\begin{equation}
\mathbf{n}_{\rm obs}'\mathbf{n}_{\rm ph}'=1-(1-\mathbf{n}_{\rm obs}\mathbf{n}_{\rm ph})\delta{\cal D}_*\,.
\end{equation}
\end{itemize}
\end{appendix}
\end{document}